\documentclass[aps,prm,reprint,superscriptaddress]{revtex4-1}

\usepackage{graphicx}
\usepackage{dcolumn}
\usepackage{epstopdf}
\usepackage{multirow}
\usepackage{color}

\newcolumntype{d}[1]{D{.}{.}{#1}}
\begin{document}

\title[Vibrational properties of monolayer TMDs]  {Vibrational and dielectric properties of monolayer transition metal dichalcogenides}

\author{Nicholas A. Pike}
\email[]{Nicholas.pike@smn.uio.no}
\affiliation{Centre for Materials and Nanotechnology, University of Oslo, NO-0349 Oslo, Norway}
\affiliation{Nanomat/Q-Mat/CESAM, Universit\'{e} de Li\`{e}ge \& European Theoretical Spectroscopy Facility, B-4000 Li\`{e}ge, Belgium}
\author{Antoine Dewandre}
\affiliation{Nanomat/Q-Mat/CESAM, Universit\'{e} de Li\`{e}ge \& European Theoretical Spectroscopy Facility, B-4000 Li\`{e}ge, Belgium}
\author{Benoit Van Troeye} 
\affiliation{Universit\'{e} catholique de Louvain, Institute of Condensed Matter and Nanosciences (IMCN) \& European Theoretical Spectroscopy Facility, B-1348 Louvain-la-Neuve, Belgium}
\author{Xavier Gonze}
\affiliation{Universit\'{e} catholique de Louvain, Institute of Condensed Matter and Nanosciences (IMCN) \& European Theoretical Spectroscopy Facility, B-1348 Louvain-la-Neuve, Belgium}
\author{Matthieu J. Verstraete}
\affiliation{Nanomat/Q-Mat/CESAM, Universit\'{e} de Li\`{e}ge \& European Theoretical Spectroscopy Facility, B-4000 Li\`{e}ge, Belgium}

\date{\today}
\begin{abstract}

First-principles studies of two-dimensional transition metal dichalcogenides have contributed considerably to the understanding of their dielectric, optical, elastic, and vibrational properties. The majority of works to date focus on a single material or physical property. Here we use a single first-principles methodology on the whole family of systems, to investigate in depth the relationships between different physical properties, the underlying symmetry and the composition of these materials, and observe trends. We compare to bulk counterparts to show strong interlayer effects in triclinic compounds. Previously unobserved relationships between these monolayer compounds become apparent. These trends can then be exploited by the materials science, nanoscience, and chemistry communities to better design devices and heterostructures for specific functionalities.

\end{abstract}
\pacs{Transition metal dichalcogenides, dielectric properties, vibrational properties, Raman spectra, Infrared spectra, mechanical properties}

\maketitle

Since the discovery of graphene~\cite{Novoselov2004}, the search for useful two-dimensional materials~\cite{Gupta2015, Butler2013, Dubertret2015} as components of more complex electronic~\cite{Rasmussen2015, Roldan2014, Kuc2015, Kireev2018} or electro-optical devices~\cite{Alyoruk2015, Wang2012} has expanded significantly. Many two-dimensional materials display unique properties, and one can combine the individual layers~\cite{Huo2015, Jariwala2014, Radisavjevic2011, Hamm2013, Zan2017} as building blocks to produce increasingly complex devices~\cite{Novoselov2016, Ross2017}. Recent studies have demonstrated the power of density functional theory (DFT)~\cite{Martin2004} in identifying novel two-dimensional materials that have useful electronic and optical properties~\cite{2017_mounet_2D, 2017_choudhary_scirep, Haastrup2018, Li2018} and density functional perturbation theory (DFPT) in determining the stability and vibrational properties of these materials~\cite{Petretto2018, Petretto2018b}. 

There is a significant amount of work, both experimental~\cite{Chen2017, Late2014, Lezama2015, Ruppert2014, Tongay2014, Feng2015, Manas2016} and theoretical~\cite{Molina2011, Molina2015, Kumar2014, Zhao2017, Glebko2018, Guo2014}, focusing on how the properties of transition metal dichalcogenide (TMD) monolayers differ from their bulk counterparts. However, even with this copious amount of experimental and theoretical work, no wide-ranging analysis of the vibrational and dielectric properties of the monolayer compounds, and no systematic comparisons of these properties with their bulk counterparts, exist. Such an analysis is essential to understand and appreciate the unique properties that the monolayers and their bulk counterparts have. 

In these layered materials, the interlayer interaction is generally weak compared to the intralayer one~\cite{Mannebach2017, Straub1985}. Consequently, one expects the intrinsic chemistry of the constituent layers to remain relatively unaffected by stacking. To some extent, the physical properties of the bulk or van der Waals (vdW) heterostructures should be related to the properties of the individual layers. Failures of this approximation include the recent discovery of superconductivity in~\textit{bilayer} graphene twisted at a specific angle~\cite{cao_2018_supercond_graphene_bilayer}, superconductivity in bilayer graphene/h-BN layers~\cite{Moriyama2019}, and the ultralow thermal conductivity in disordered, layered, WSe$_2$~\cite{Chiritescu2007}. Additionally, when vdW layers are stacked during device construction, one or both of the lattices may compensate by expanding or contracting, to maximize the interlayer interactions at the expense of elastic energies~\cite{Woods2014}. This can result in the formation of a moir\'{e} pattern, and possible associated modification of electronic properties~\cite{Espejo2013, Lim2014, Rigosi2015, Novoselov2016, Wilson2017,  Troeye2018}, as observed for example in MoS$_2$-WSe$_2$ vdW-heterostructure~\cite{Zhang2017}. Knowledge of the structural properties of the monolayers leads to better device construction and orientation by taking into account the nature of each material (e.g. elasticity, dielectric response...).

\begin{figure*}[t]
\centering
\vspace{-0.5cm}
\includegraphics[width=0.90\textwidth]{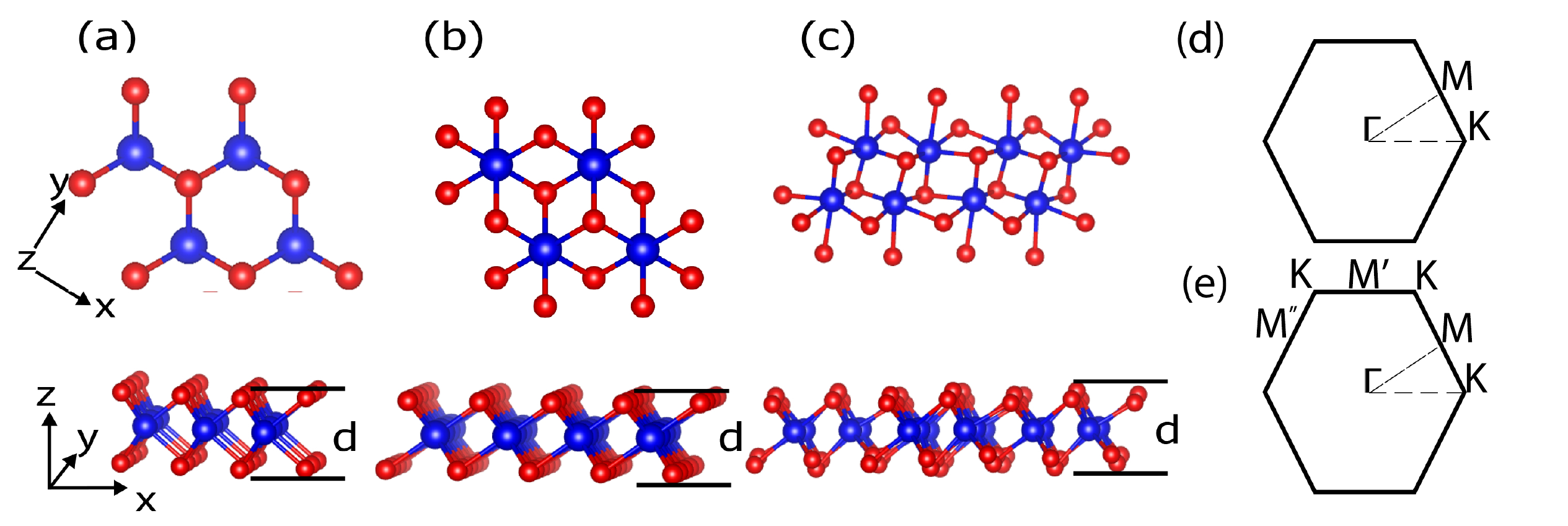}
\caption{(Color online) Sketch of a generic monolayer (a) h-TMD, (b) t-TMD, and (c) tc-TMD where the transition metal is given in blue and the chalcogen atoms are in red. (d) The Brillouin zone for the h and t compounds in two dimensions and (e) the Brillouin zone for the tc compounds in two dimensions.  }
\label{fig:unitcell}
\end{figure*}
 
The theoretically-predicted vibrational properties can easily be compared to Raman and Infrared spectroscopy experiments~\cite{Zhang2016b,Corro2016} in order to identify the number of layers, strain states, and certain defect states. The related dielectric properties are critical as they relate directly to the electro-optical properties of these materials. With this in mind, we investigate systematically the physical, electrical, dielectric, and optical properties for the most common and stable TMD monolayers. These include the in-plane lattice parameters, the thickness of an individual layer, electronic energy gap, binding energy, and the elastic, dielectric, Born effective charge, piezoelectric, and non-linear optical tensors.

In this article, we focus on the most common and stable TMDs, which cover the three different symmetry classes: hexagonal (h) -TMDs: MoS$_2$, MoSe$_2$, MoTe$_2$, WS$_2$, and WSe$_2$; trigonal (t) -TMDs:  TiS$_2$, TiSe$_2$, TiTe$_2$, ZrS$_2$, and ZrSe$_2$; and triclinic (tc)-TMDs: ReS$_2$, ReSe$_2$, and TcS$_2$. This allows us to determine the effects of chemical composition and the local environment. We find both expected trends and new ones in the elastic, dielectric and vibrational properties. In particular, in the triclinic compounds, Young's modulus increases with the mass of the transition metal, which is counter intuitive. Furthermore, some of these compounds are only 2-3 times stiffer than graphene, and thus much less so than MoS$_2$, which had previously led the community to believe that TMDs were unadapted to flexible electronics. In the main text below, a single sulfide member from each structural family is given as an example alongside the aggregate trends. Details on the remaining materials can be found in the supplemental materials (SM)~\cite{supmat1}.

\section{Calculation Methods}\label{calculation}

To determine the properties of our chosen materials, DFT~\cite{Martin2004, Gonze1997b} and DFPT~\cite{Baroni2001, Gonze1997a, verstraete_2014_dfpt} calculations are undertaken using the~\textsc{Abinit} software package~\cite{Gonze2005, Gonze2009, Gonze2016}. Consideration of the usual DFT approximations (pseudopotential, exchange-correlation, dispersion correction, and spin-orbit coupling) is summarized in Ref.~\cite{bulk_pike} for the bulk counterparts of these systems, and are systematically checked during our investigations of the monolayers. 

For our calculations, we have used the GGA-PBE exchange-correlation functional and Trouiller-Martins pseudopotentials~\cite{Martin2004, Perdew1996, Fuchs1999}, generated with the fhi98pp code for all elements except W and Ti. In the case of W, we exploit a pseudopotential generated with the~\textsc{Opium} code~\cite{opium_psp}, which produces accurate relaxed lattice parameters, while for Ti, we use an ONCVPSP~\cite{Hamann2013} generated pseudo-potential, which correctly reproduces the Kohn anomalies in the corresponding materials. Long-range dispersion forces are included using Grimme's DFT-D3 scheme~\cite{Grimme2010}, as implemented in~\textsc{Abinit}, for both the ground-state and response-function parts of the code~\cite{Troeye2016, Troeye2017}. All electronic and response-function computations use the relaxed geometries. The residuals for the ground-state and first-order wave-function were converged below 10$^{-18}$ and 10$^{-10}$, respectively. We perform convergence studies of the energy cut-off and the reciprocal space $k$ sampling using a Monkhorst-Pack grid~\cite{Monkhorst1976} such that the total energy change was less than 0.01 meV. These studies provide a range of energy cut-offs between 20 and 50 Ha, depending on the atomic species as outlined in the SM~\cite{supmat1}, and a k-point mesh of $8\times8\times1$. During structural relaxation, we use the Broyden-Fletcher-Goldfarb-Shanno minimization procedure~\cite{Gonze2009} to relax the positions and unit cell simultaneously, with a maximum force below $1.0\times10^{-7}$ Ha/Bohr.

Our calculated Raman spectra and nonlinear optical tensor~\cite{Veithen2005}, use LDA exchange-correlation for calculation of the third derivative of the exchange-correlation part of the energy~\cite{Ceperley1980}. We have rescaled the Raman intensities by their maximum value: usually experiments do not report absolute values and only relative differences in intensity are comparable. In each case, we plot the Raman frequencies assuming a fixed Lorentz broadening for the spectral lines of $1.0\times10^{-5}$ Ha and a laser wavelength of 532 nm, a typical frequency used in Raman experiments~\cite{Zeng2013, Wolverson2016}.  Our plotted Raman data include in-plane (XX and YY), out-of-plane (ZZ), and powder averaged spectra, which are calculated assuming a random orientation of monolayer flakes, similar to what is done in Ref.~\citenum{Caracas2006}.

\begin{table*}[!th]
\caption{Properties of monolayer MoS$_2$, ZrS$_2$ and ReS$_2$ with our calculations in the first sub-column, literature values in the second, and the reference in the third. We report the in-plane lattice parameters, $a$, and $b$, the geometric thickness of an individual layer, $d$, the in-plane components of the elastic tensor per unit area ($c_{ij}$), Young\rq{}s modulus ($E_i$), Poisson's ratio ($\nu_{ij}$), bending rigidity ($\kappa_i$), Kohn-Sham electronic band gap energy ($E_g$,) binding energy ($E_b$), in-plane components of the dielectric tensor ($\epsilon^0_{ij}$), optical dielectric tensor ($\epsilon^\infty_{ij}$), and Born effective charges ($Z^*_{ii}$) on the transition metal atom (blue in Fig.~\ref{fig:unitcell}), piezoelectric coefficient ($e_{11}$), non-linear optical coefficient ($d_{16}$), Debye temperature ($\theta_D$), average speed of sound (v$_{avg}$), and the Helmholtz Free-energy at zero temperature, ($\Delta F(0)$). $^*$ indicates the predicted gap is indirect, $^\dagger$ indicates the values are divided by 2 for the sake of comparison with other TMDs,  f.u. = formula unit and values in brackets correspond to other first-principles calculations. For the tc-compounds, the two lines of the Born effective charge tensor correspond to the two in-equivalent transition metal atoms.}

\label{Tab:props}
\centering
\small
\begin{tabular}{|r|d{3.3} d{7} l |d{3.4} d{7} l | d{3.4} d{7} l | }\hline
& \multicolumn{3}{c|}{MoS$_2$} & \multicolumn{3}{c|}{ZrS$_2$}& \multicolumn{3}{c|}{ReS$_2$}\\
& \multicolumn{1}{c}{Calc.}& \multicolumn{1}{c}{Lit.}&\multicolumn{1}{c|}{Ref.}& \multicolumn{1}{c}{Calc.} & \multicolumn{1}{c}{Lit.} &\multicolumn{1}{c|}{Ref.}& \multicolumn{1}{c}{Calc.} & \multicolumn{1}{c}{Lit.}&\multicolumn{1}{c|}{Ref.} \\ \hline

a ({\AA}) & 3.165 & 3.200 [3.19]& \cite{Joensen1987,2017_mounet_2D} & 3.695 &[3.68, 3.67]&\cite{Guo2014,2017_mounet_2D} &6.414 & & \\
b ({\AA}) & & & & &&&6.581 & & \\
d ({\AA}) & 3.186 & 3.172& \cite{Joensen1987} & 2.967 &3.625^\dagger&\cite{Wang2016}& 3.541 &3.50^\dagger&\cite{Tongay2014} \\

$E_g$ (eV) & 1.896 & 1.85 [1.6]& \cite{Huo2015} & 1.449^* &[1.02^*,1.2^*]&\cite{Zhao2017,2017_mounet_2D} &1.442^* & [1.85^*]&\cite{Gehlmann2017}\\
$E_b$ (meV/$\AA^2$) & 30.316& [21.6;28.8] &\cite{2017_mounet_2D} & 19.043&[19.0;24.1]&\cite{2017_mounet_2D} &1.010 && \\ \hline

c$_{11}$ (Ha/Bohr$^2$) & 0.082 && & 0.042&& &0.106 && \\
c$_{22}$ (Ha/Bohr$^2$) & & & & && &0.108 && \\
c$_{12}$ (Ha/Bohr$^2$) & 0.018 && & 0.006&&&0.019 && \\
c$_{66}$ (Ha/Bohr$^2$) & 0.032 && & 0.018&&& 0.004&& \\
E$_x$ (Ha/Bohr$^2$) & 0.078 & 0.110 & \cite{Liu2014} & 0.041 &&& 0.102&&\\
E$_y$ (Ha/Bohr$^2$) & & & &&&& 0.104 &&\\
$\nu_{xy}$  & 0.218 &[0.25] &\cite{Kang2013}& 0.14 &&&0.182 &[0.207]& \cite{flexibZhao2017a}\\
$\kappa_x$ (eV)& 10.96 & 9.93 &\cite{Casillas2015} & 5.17 &&&3.09 && \\

$\epsilon^0_{xx}$ & 20.20 && &43.35 & &&23.28 && \\
$\epsilon^0_{yy}$ & & & & && &20.85 && \\
$\epsilon^0_{xy}$ & & & && &&0.15 && \\
$\epsilon_{xx}^\infty$ & 19.98& & & 14.48&&& 18.72 && \\
$\epsilon_{yy}^\infty$ & & & &&&& 20.65 && \\
$\epsilon_{xy}^\infty$ & & & &&&& 0.25 && \\
$Z_{xx}^*$ (e) & -1.088 & [-1.004]&\cite{Sohier2017b}&6.205 &&&-1.496 && \\
& & & &&&& -0.532 && \\
$Z_{yy}^*$ (e) & & & &&&& -0.556 && \\
& & & &&&& 0.296&& \\
e$_{11}$ ($\times 10^{-10}$C/m)&2.77 &2.9& \cite{Zhu2015a} & & && && \\
d$_{16}$ ({\AA}$^2$/V)& 0.406& 1.0 &\cite{Woodward2017} & & &&& & \\ \hline 
$\theta_D$ (K) & 177.48 & && 133.08 &&&86.72 && \\
v$_{avg}$ (km/s) & 4.068&& & 3.269 & [3.4]&\cite{Glebko2018}&3.213 && \\ 
$\Delta F(0) $(kJ/mol) & 14.567& && 10.962&&& 46.093&&\\ \hline
\end{tabular}
\end{table*}

\section{Results and Discussion}
All TMDs consist of a layer of metal atoms packed between two layers of chalcogen atoms. The monolayer h-TMDs show 6-fold octahedral coordination (in the form of a trigonal prism) in the layer and space group $P\overline{6} m2$. In Fig.~\ref{fig:unitcell}(a) we show a typical h-TMD layer in the $xy$ and $xz$ planes containing three atoms per unit cell. The t-TMDs consist of chalcogen atoms forming a trigonal anti-prism, with space group $P\overline{3}m1$, as shown in Fig.~\ref{fig:unitcell}(b) where there are two chalcogens and one transition metal atom per unit cell.  Finally, in Fig.~\ref{fig:unitcell}(c) we show a typical tc-TMD composed of 12 atoms, grouped into two types of in-equivalent distorted octahedra, and $P\overline{1}$ space group symmetry. Several authors~\cite{Gehlmann2017, Hart2017, Webb2017} have noted that the two-dimensional Brillouin zone of these materials is similar to the in-plane Brillouin zone of a hexagonal or trigonal system, as shown in Fig.~\ref{fig:unitcell}(e). There are inequivalent M points on the faces of the Brillouin zone, due to the difference in lattice vector lengths. In the band structures and Raman spectra that follows for the tc compounds, we plot only the path $\Gamma$-M-K-$\Gamma$, where M corresponds to ($\frac{1}{2}$,0,0), the a axes of the h, t,  and tc compounds have been aligned, and we use 120 degree unit cells for h- and t- systems.

The accurate calculation of monolayer properties with periodic boundary conditions requires a large vacuum to separate the periodic images.  To compare the material properties which depend on volume, such as the dielectric tensor, to their bulk counterparts, the unit cell volume must be rescaled. Similarly, the reduced dimensionality of the system means that the phonon band structures, and corresponding density of states, have different frequency dependencies in two- and three-dimensions. This dimensionality manifests itself in a distinct quadratic acoustic mode in the phonon band structure and as an apparent logarithmic divergence in the thermal properties of the system. In reality, the divergences of the thermal properties are suppressed by this quadratic ZA mode which plays a dual role as a significant source of occupied phonon states and as a scattering source as shown for graphene~\cite{Lindsay2010, Pereira2013}. We do not present thermal properties here as our calculations stay at the harmonic level in~\textsc{Abinit}.

\begin{figure*}[!t]
\centering
\includegraphics[width=0.25\textwidth]{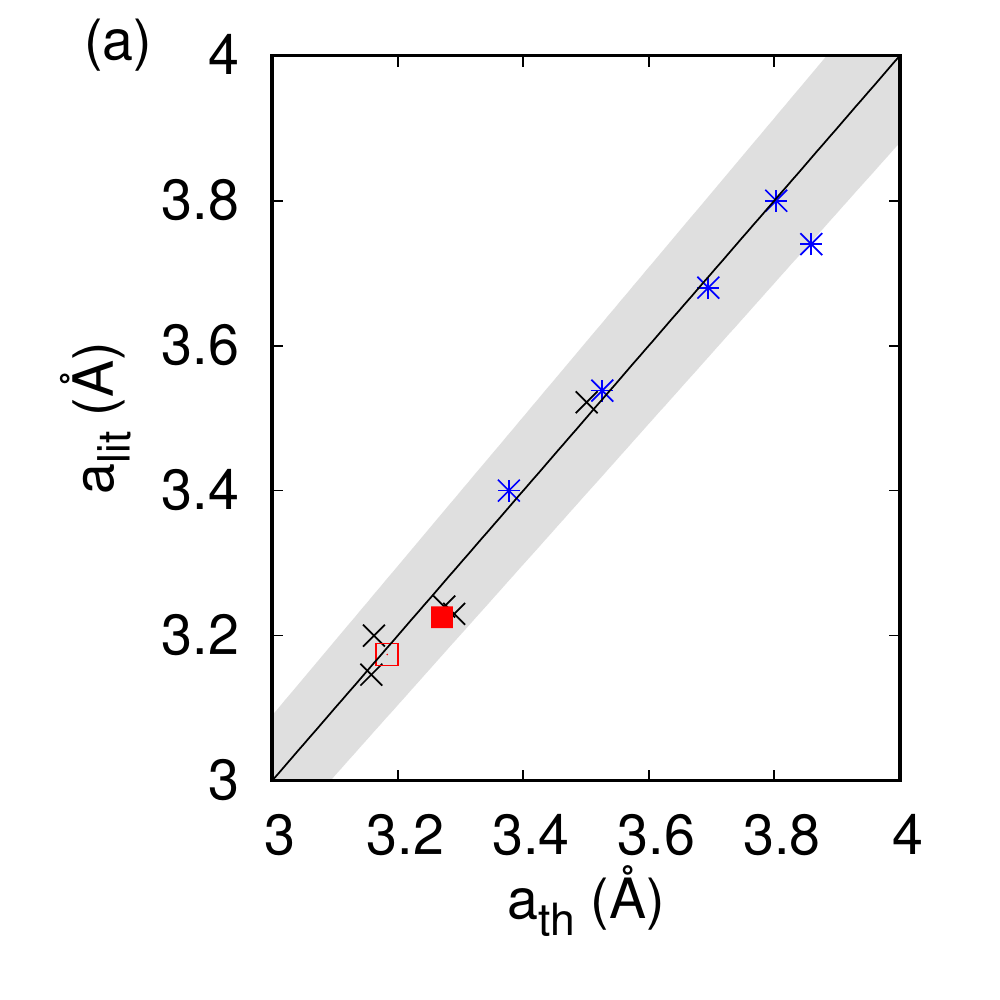}
\includegraphics[width=0.25\textwidth]{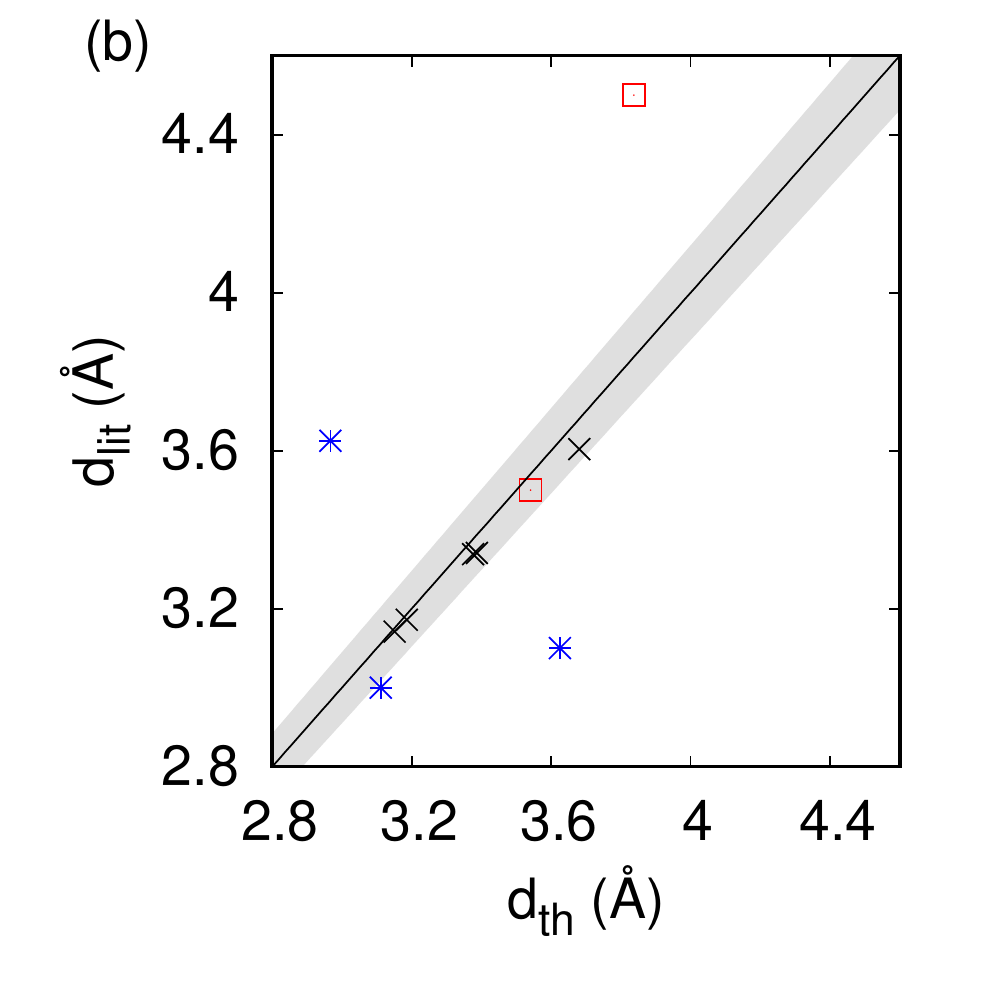}
\includegraphics[width=0.25\textwidth]{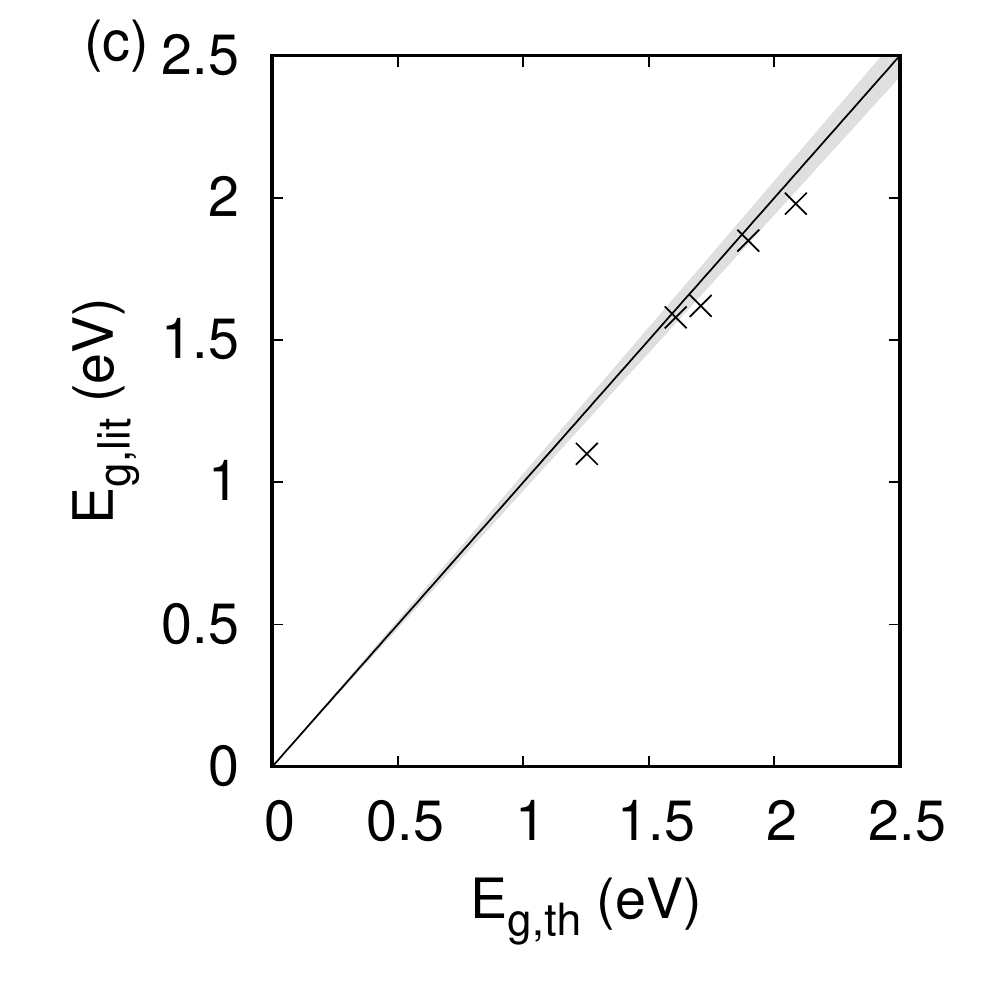}\\
\includegraphics[width=0.25\textwidth]{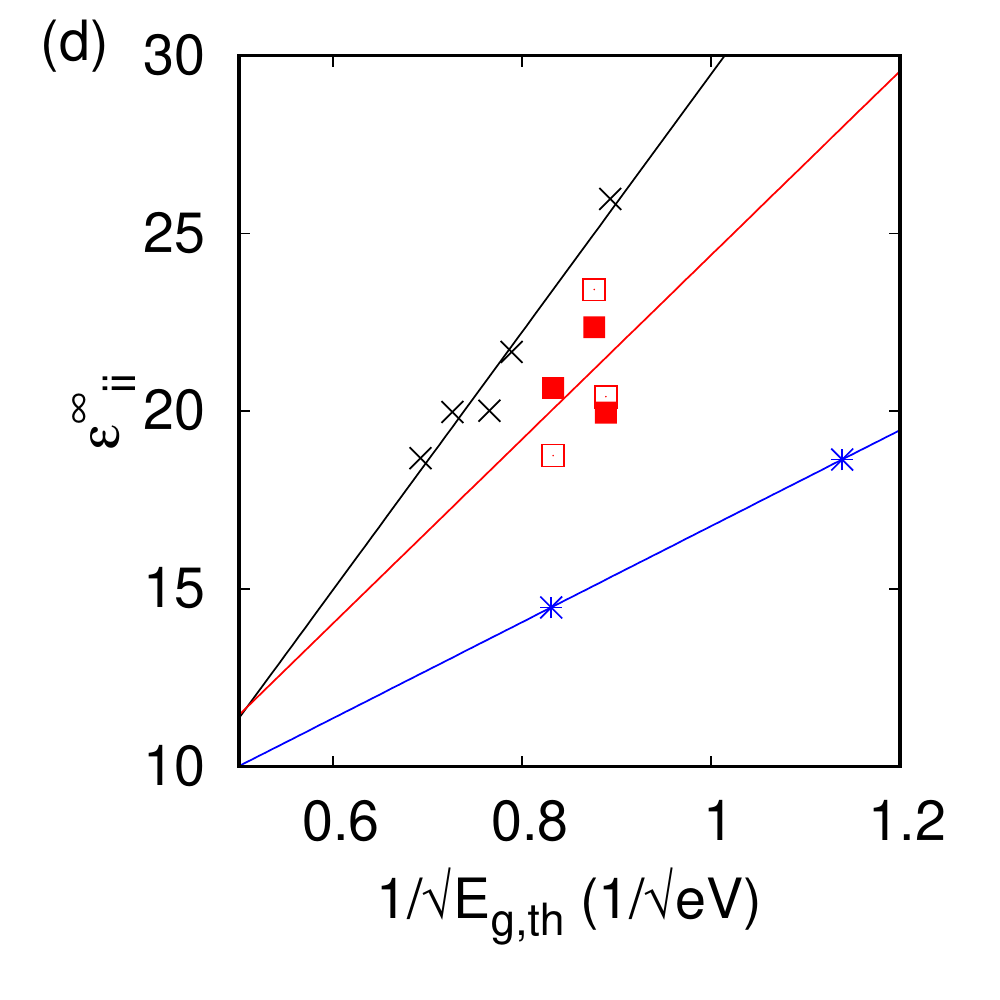}
\includegraphics[width=0.25\textwidth]{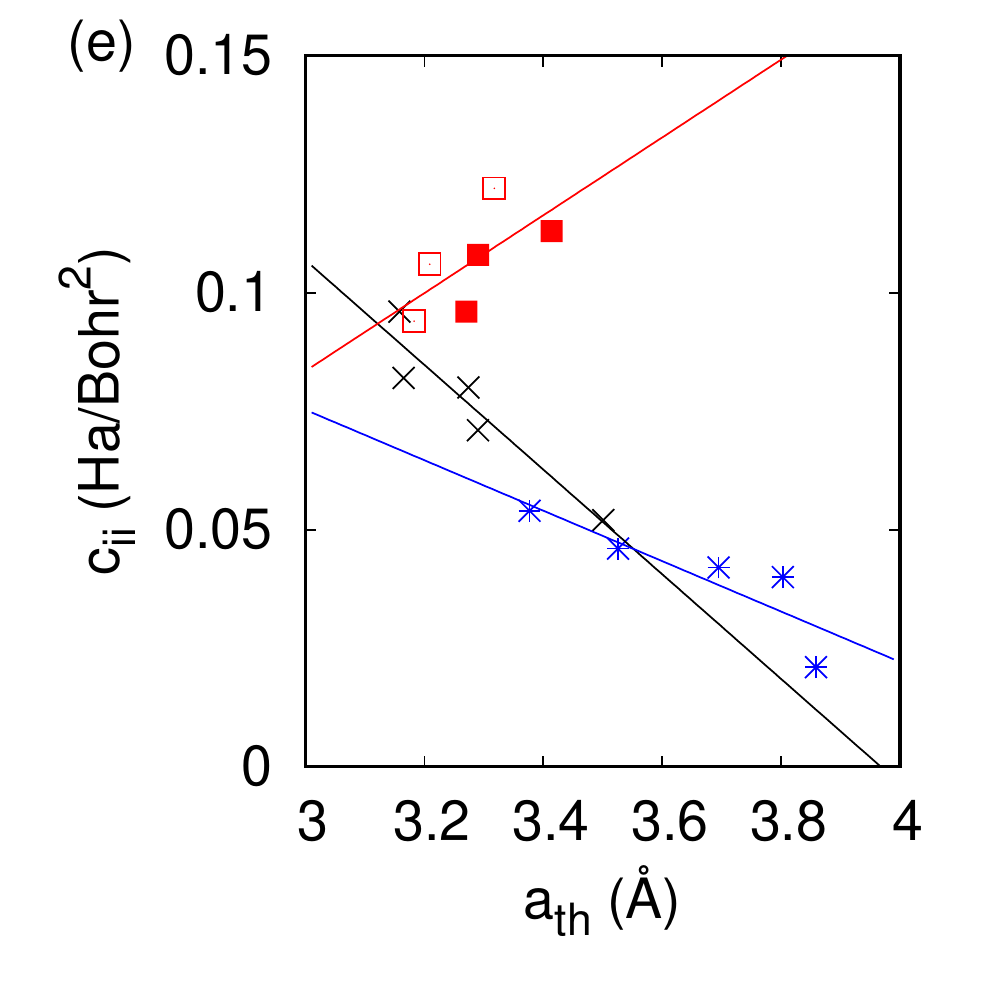}
\includegraphics[width=0.25\textwidth]{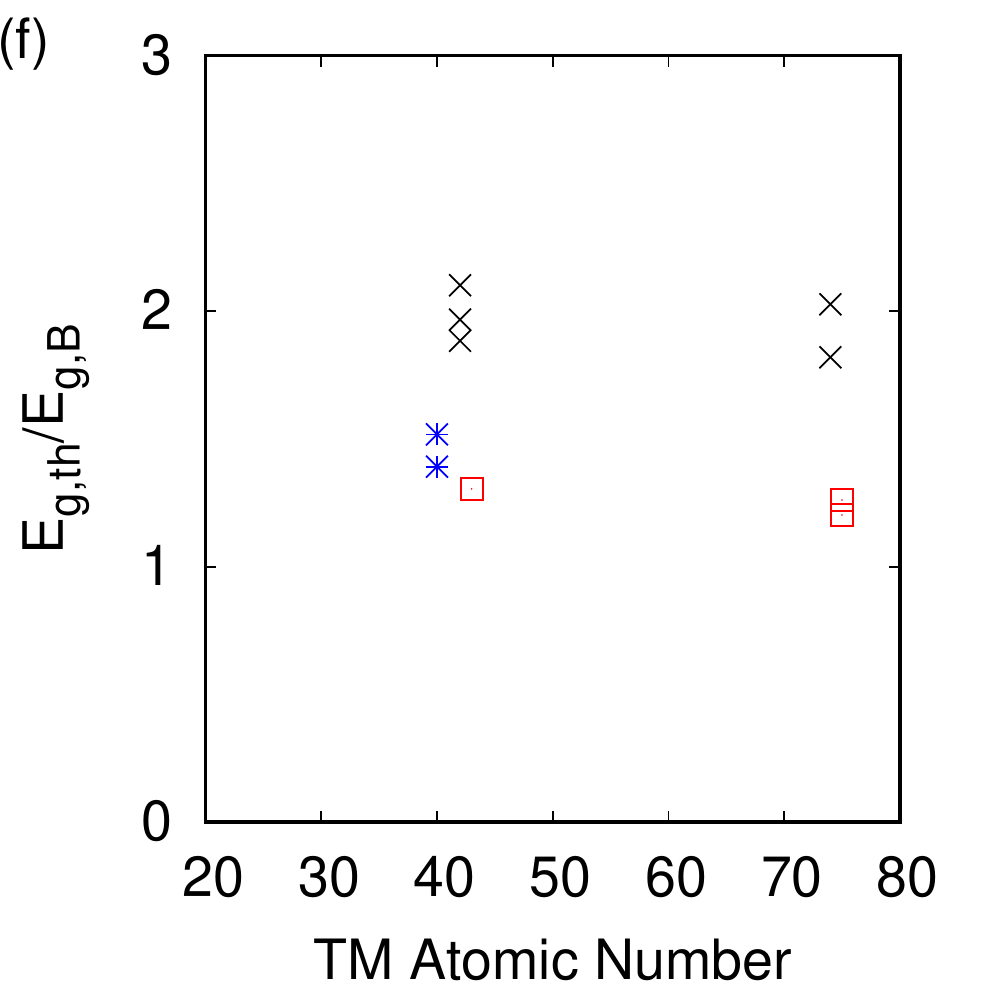}\\
\includegraphics[width=0.25\textwidth]{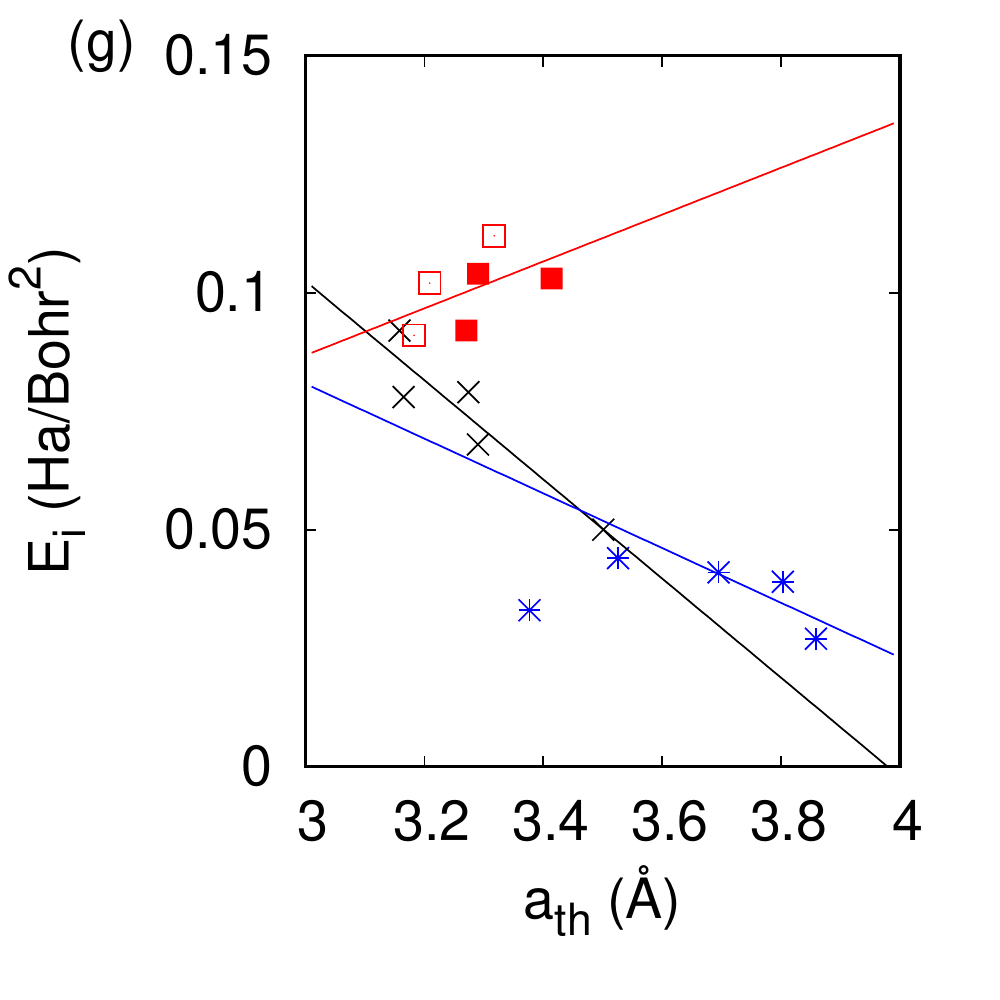}
\includegraphics[width=0.25\textwidth]{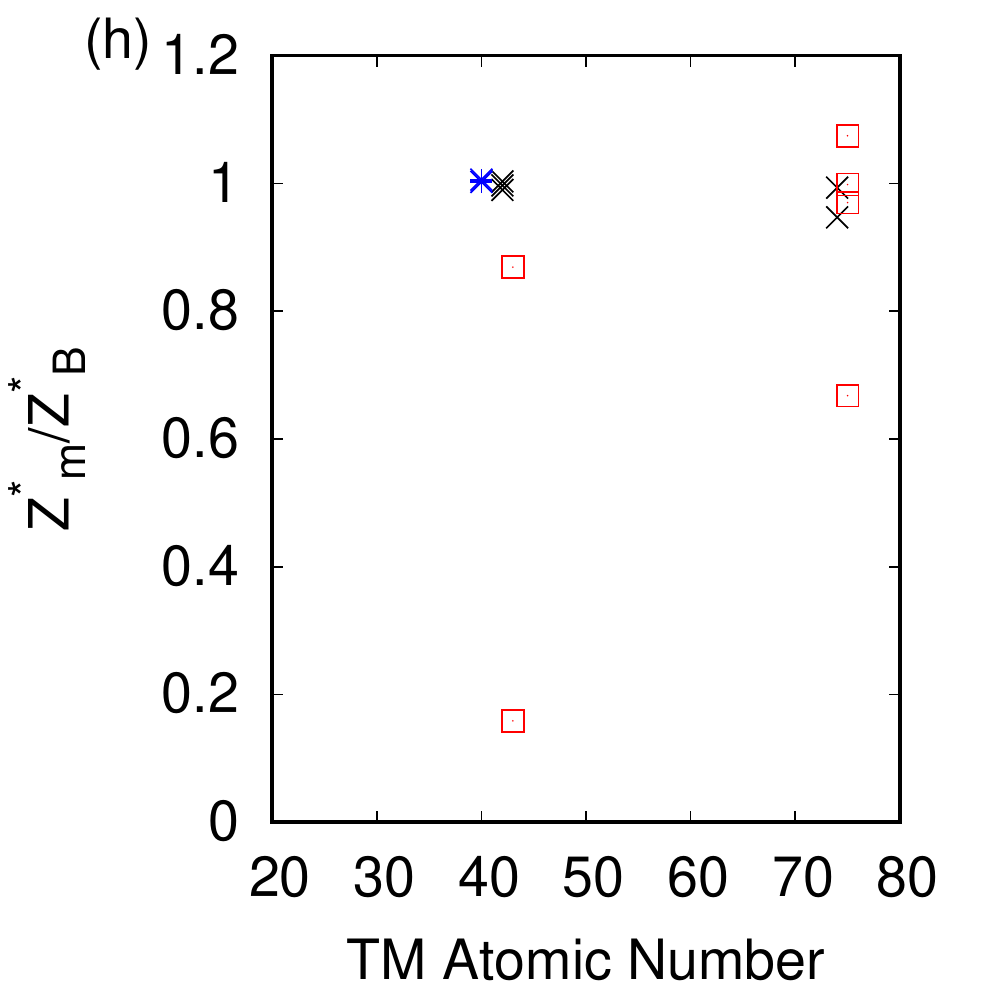}
\caption{(Color Online) Comparison between our theoretical values (th)  and literature values (lit) for the principal dielectric and structural properties of TMDs. h-TMDs are shown as black crosses, t-TMDs as blue stars, and tc-TMDs as red squares (xx component as empty and yy component as solid squares). (a) the in-plane lattice parameter, (b) thickness of the monolayer, and (c) band gap energy for the monolayers. The solid gray wedge in (a) - (c) represents $\pm$3\% error. In (d) we show the relationship between the dielectric response and the inverse square root of the calculated energy gap, with the solid line indicating a linear fit of the semiconducting TMDs for each family. In (e) we show the relationship between the in-plane elastic constants and the in-plane lattice parameters. (f) and (h) compare monolayers with bulk, for calculated band gap energies and Born effective charges. Finally, in (g) the calculated Young's modulus vs lattice parameter is plotted.  In (a), (e) and (g)  the lattice parameter of the tc compounds is divided by two for the sake of comparison with other TMDs.}
\label{trends}
\end{figure*}

In Table~\ref{Tab:props} and Fig.~\ref{trends}(a) and (b), we show our calculated in-plane lattice parameters with the corresponding available literature data~\cite{Joensen1987, Boker2001, Chang2014, Berkdemir2013, Wang2015, Han2016, Zeng2011, Chen2015, Guo2014}. The lattice parameters should be compared to their bulk counterparts (MoS$_2$: 3.162 \AA, ZrS$_2$: 3.687 \AA, ReS$_2$: 6.420 and 6.587 \AA)~\cite{bulk_pike} and show only slight changes (0.2\%) of the in-plane lattice parameters in Fig.~\ref{trends}(f). For the tc compounds, the calculated angle between the a and b lattice vectors is 60.2 degrees.  The thickness of an individual layer, d, defined here as the vertical distance between the outermost chalcogen atoms, compares favorably to experimental data on free-standing layers for the h compounds, and gives a reasonable comparison to half the experimentally measured thickness in measurements of monolayers on substrates~\cite{Wang2016, Peng2015, Mleczko2017, Jiang2018, Tongay2014}. Note that the observed differences between our calculated thickness and half the experimental thickness are most likely due to the changing interaction between the monolayer and substrate. 

The binding energy E$_b$ of a given TMD material is defined as the gain of energy resulting from the stacking of the corresponding TMD monolayers on top of each other in the optimal stacking sequence. In other words, E$_b$ represents the difference in energy between the total energy of the bulk TMD compound (E$_{T,bulk}$) and the number of layers per primitive cell (n) times the total energy of a monolayer (E$_{T,mono}$) as
\begin{equation}
E_{b} = E_{T,bulk} -nE_{T,mono}.
\end{equation}
While most of the binding energies for these TMD monolayers lie between 20 and 40  meV/\AA$^2$, which is consistent with Ref.~\citenum{2017_mounet_2D}, the tc-TMD compounds have binding energies on the order of only 1 meV/\AA$^2$. This small value is consistent with experimental Raman measurements and DFT interfacial interaction energies, indicating that the interlayer coupling in the tc compounds is relatively weak~\cite{Tongay2014, Feng2015, Zhao2015a}. In Raman measurements there is only a very small shift in the peak frequencies between the monolayer and the bulk, and the ZO optical mode in tc bulk is lower than in MoS$_2$~\cite{Tongay2014}.

The planar elastic constants (per unit area - Table~\ref{Tab:props}) are comparable to their bulk counterparts, when rescaled to account for the vacuum spacing. In two dimensions, the only non-zero components of the elastic tensor are $c_{11}$, $c_{12}$, $c_{22}$ and $c_{66}$ ($c_{66} = (c_{11}-c_{12})/2$ for the hexagonal and trigonal systems)~\cite{Nye1957, Tschoegl1957}. A direct comparison between these elastic tensor components and their bulk values reveals almost no difference, as is often assumed in finite element analysis~\cite{Dluzewski2010}. Our DFT calculations are an important sanity check for the whole set of common TMD.

To compare to experiment, we calculate several derived properties such as the Bulk modulus, Young's modulus, and Poisson's ratio, taking into account the two-dimensional nature of the system as shown in the SM~\cite{Tromans2011, Nye1957, supmat1}. Our calculated Young's modulus agrees well with experimental measurements on free-standing monolayers~\cite{Liu2014, Zhang2016}. 

Of particular interest is the relationship between our calculated in-plane lattice constant and the components of the elastic tensor for each symmetry class of compounds. As shown in Fig.~\ref{trends}(e) there exists a roughly linear relationship between the in-plane lattice parameter and c$_{11}$ over the range of compositions and chemistries. Ref.~\cite{Zhang2016a} suggests the decrease in Young's modulus of these materials is related to the increase in the lattice parameter, and the decrease in charge transfer. We have plotted this first relationship in Fig.~\ref{trends}(g) and demonstrate that the relationship with lattice parameter holds for the h- and t-TMDs but not for the tc-TMDs. The Born effective charge quantifies the charge transfer and polarizability in these monolayer systems; a comparison to their bulk counterparts is given in Fig.~\ref{trends}(h) where the ratio of the monolayer to the bulk Z$^*$ of Ref.~\cite{bulk_pike} is given. Fig.~\ref{trends}(h) reveals no difference in these systems for the h- and t- compounds, but does show a strong scatter in the Born effective charge in the tc compounds: isolating a monolayer rearranges the charge and polarizability of the inequivalent octahedra.  As a measure of the static charge rearrangement, we present the Bader charges of the tc compounds compared to their bulk values and compare these with the bulk and monolayer Bader charges of MoS$_2$ in the SM~\cite{supmat1}. We find that the Bader charge also differs between the monolayer and the bulk for the triclinic compounds, but there is no such difference for MoS$_2$.

It was recently demonstrated~\cite{Dai2016, Troeye2017} that the bending modes are of critical importance to our understanding of moir\'e patterns in vdW-heterostructures. We calculate the bending rigidity, $\kappa$, using Kirchhoff-Love theory for thin plates~\cite{Love1888} as:
\begin{equation}
\kappa = \frac{E d^3}{12(1-\nu^2)} \label{kappa}
\end{equation}
where $d$ is the thickness of the individual layer with Young's modulus in GPa. As noted in Refs.~\cite{Lai2016} and~\cite{Lindahl2012}, the definition of the thickness, $d$, of the thin plate is potentially ambiguous when dealing with two-dimensional materials. Experimentally, the thickness of these materials frequently includes the height of the vdW gap. However, as we calculate free-standing monolayers, there is no vdW gap and we define the thickness as the vertical distance between the outermost atoms. This value is easily calculated in DFT and compares well to measurements of free-standing layers~\cite{Casillas2015}. With this definition of the thickness, we agree well with the theoretical calculations of the bending rigidity of Lai {\it et al.}~\cite{Lai2016}. We calculate the bending rigidity using the larger of the two in-plane Young's moduli and the maximum thickness of the layer, as reported in Table~\ref{Tab:props}. The bending rigidity of these compounds is significantly greater than for graphene ($\kappa = 1.2$ eV~\cite{Nicklow1974}), with the notable exception of the tc-TMDs: with their increased internal degrees of freedom and different sizes of octahedra, the materials are just two to three times stiffer than graphene, making them two to three times softer than MoS$_2$, and potential candidates for flexible optoelectronics. 

The piezoelectric tensor can be non-zero, due to the lack of inversion symmetry, for the h compounds. Our calculations provide a distinct non-zero component of the piezoelectric tensor which can be compared to Zhu {\it et al.}~\cite{Zhu2015a} and other theoretical results~\cite{Sevik2017}, after taking into account the vacuum spacing. 
\begin{figure}[!t]
\centering
\vspace{-0.2in}
\includegraphics[width=0.5\textwidth]{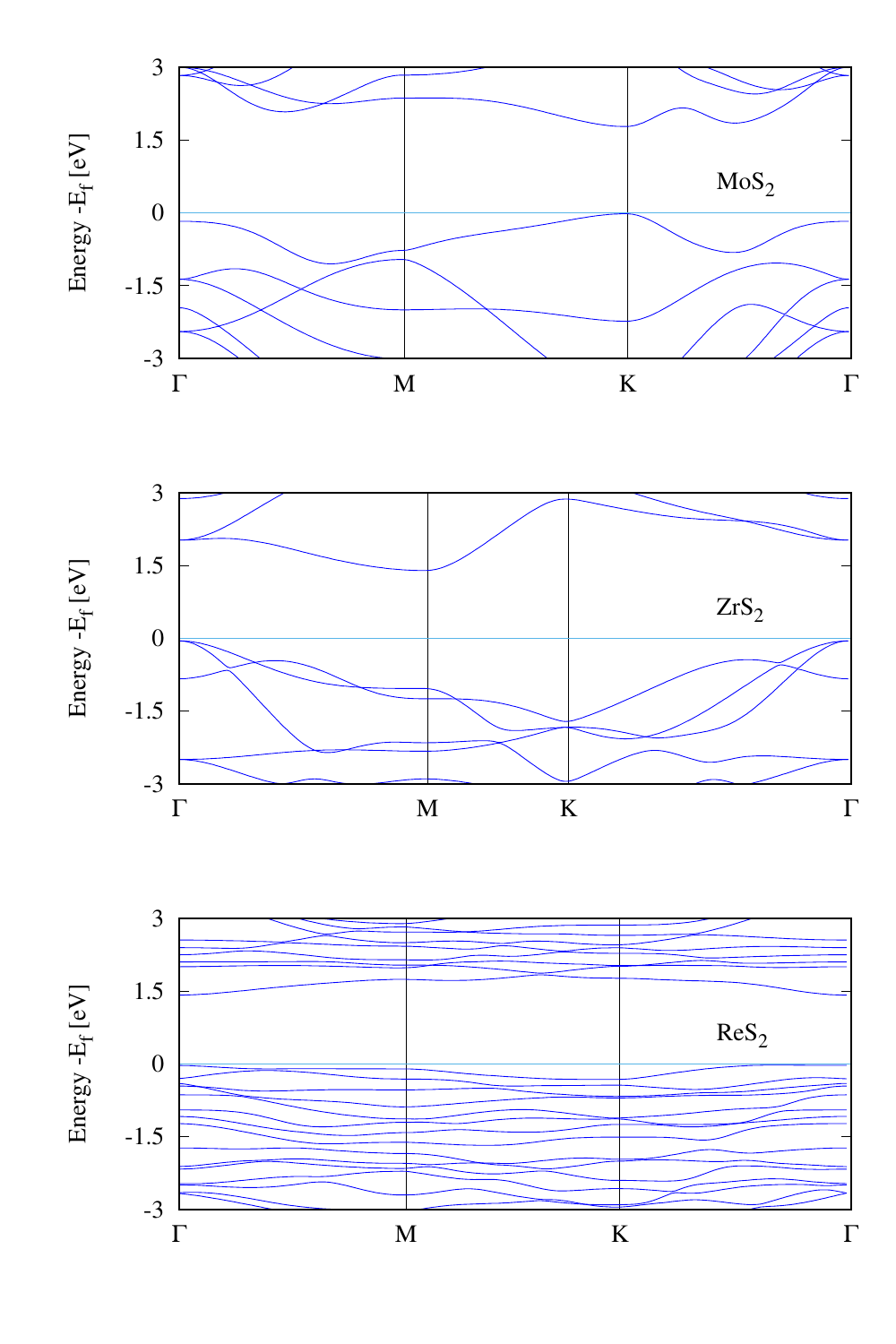}
\vspace{-0.5in}
\caption{\label{three_elec_bands} (Color Online) Calculated Kohn-Sham band structure for the three example materials with the Fermi energy shifted to the zero of energy. The band structure is plotted along a path of high symmetry in reciprocal space. }
\end{figure}

\begin{figure*}[!t]
\centering
\vspace{-0.1in}
\includegraphics[width=0.8\textwidth]{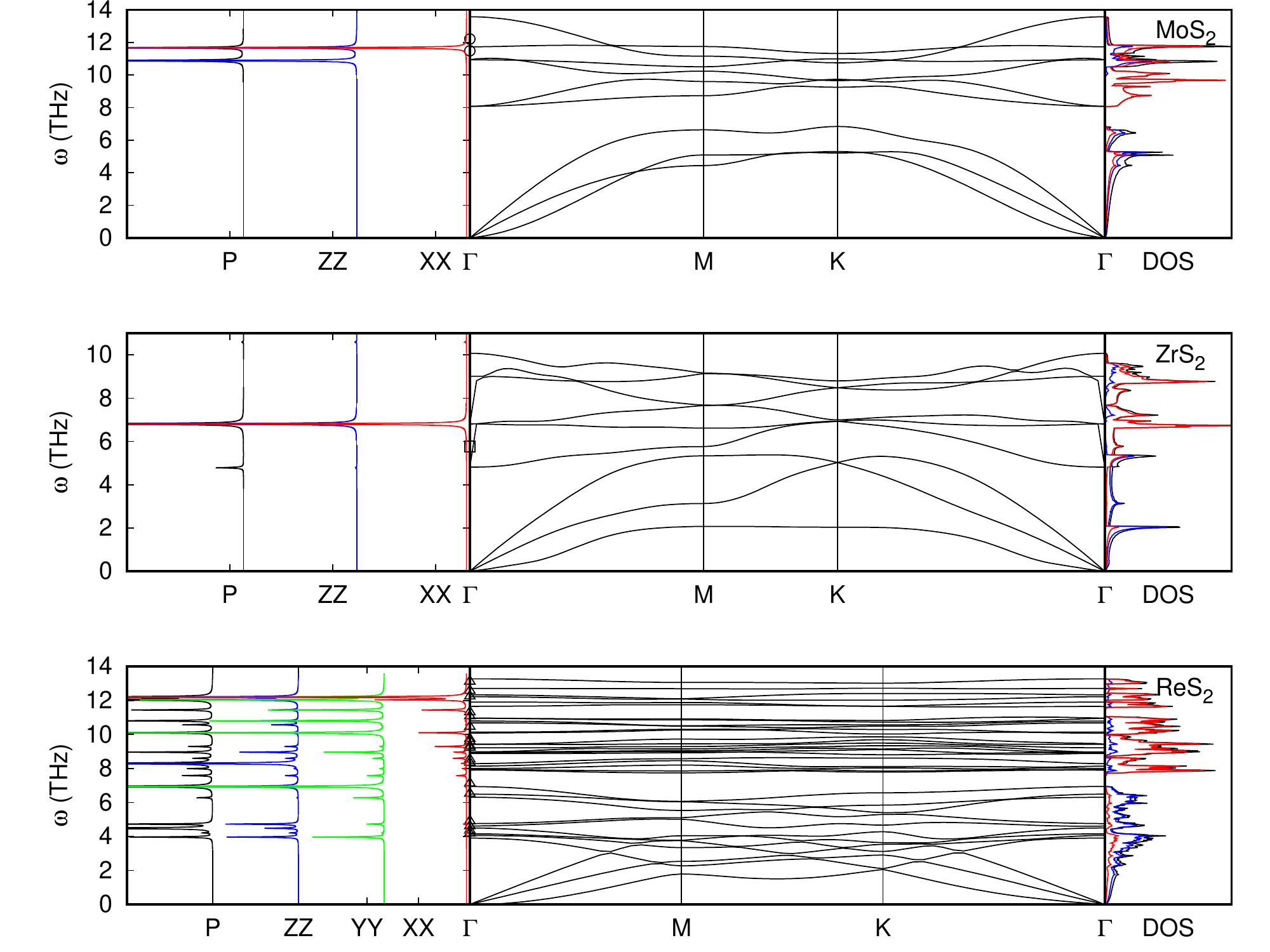}
\vspace{-0.2in}
\caption{(Color Online) Raman spectra, phonon band structure, and phonon density of states of the example materials. Experimental data from Refs.~\cite{Scheuschner2015, Feng2015, Manas2016} is shown as points at $\Gamma$. The high-symmetry path in reciprocal space is from Ref.~\cite{Setyawan2010}. For the Raman spectra, red lines correspond to XX polarization, green lines to YY polarization, and blue to ZZ polarization. The black Raman spectra correspond to a powder spectra. In the density of states plots, the blue line corresponds to the density of states of the transition metal atom, the red line corresponds to the chalcogen atom, and the black line corresponds to the total density of states. }
\label{three_phon}
\end{figure*}

The calculated Kohn-Sham electron band structures for our compounds are given in Fig.~\ref{three_elec_bands}. Monolayer h-TMDs display both direct and indirect electronic band gaps agreeing with earlier work~\cite{Ruppert2014, Hsu2017}, and which depend strongly on the residual strain in the system. The electronic band gaps are given in Table~\ref{Tab:props} and denoted with a star when they are indirect. Good agreement is found between our calculated band gap energies and experiment~\cite{Huo2015, Zhang2014, Ruppert2014, Lu2015, Gehlmann2017} for the h-TMDs as shown in Fig.~\ref{trends}(c). Some of the t-TMDs are metallic while the tc-TMDs are all semiconducting. The opening of the electronic band gap when compared to the bulk is given in Fig.~\ref{trends}(f) showing the ratio of the calculated band gap of the monolayer, $E_{g,m}$, to that for the bulk, $E_{g, B}$ (from Ref.~\cite{bulk_pike}) regardless of whether the band gaps are direct or indirect. For all the non-metallic materials, we find an opening of the electronic band gap. For the hexagonal materials, this is approximately a doubling compared to their bulk counterparts and agrees with recent work indicating that, as the number of layers decreases, the nature of the dielectric environment can change considerably (e.g. Ref.~\cite{Kang2016}).

Although DFT normally underestimates the bandgap of bulk materials, it produces relatively reasonable agreement with absorption experiments for the h-TMDs, due to the confined nature of the exciton~\cite{Hill2015, Saigal2016, Rigosi2016, Handbicki2015, Kylanpaa2015} that nearly perfectly counterbalances the intrinsic DFT Kohn-Sham band gap error~\cite{Aryasetiawan1998, Perdew1985}. The nearly degenerate conduction band minima found in the h-TMDs indicate that a small amount of strain can easily change these materials from a direct gap to an indirect gap semiconductor~\cite{Lezama2015, Wang2015a}. 

To compare our calculated dielectric tensors to their bulk counterparts, we must once again take into account the vacuum spacing. A comparison between $\epsilon^\infty$ and the inverse square root of our calculated band gap energy in Fig.~\ref{trends}(d) shows the expected linear dependence of these two quantities~\cite{Czaja1963} with different proportionality constants for each symmetry class. The magnitude of the Born effective charge tensor in these materials can be experimentally measured in Infrared reflectivity experiments. The sign of the Born effective charge and its origin are discussed extensively in Ref.~\cite{Pike2016} and are counterintuitive for the h- and tc-TMDs. The Born effective charges calculated here agree with previous DFT and DFPT calculations~\cite{Sohier2016, Danovish2017, Sohier2017b}, and with their bulk counterparts as shown in Fig.~\ref{trends}(h). While no significant changes are observed for the h- and t- compounds, the magnitude (and sign for TcS$_2$) of the Born effective charge changes for the tc- compounds due to differences in the charge and polarizability between the monolayer and the bulk. 

We compare our nonlinear optical tensor to experimental work~\cite{Woodward2017, Seyler2015} after rescaling to account for the vacuum spacing, as was done in Ref.~\cite{Wang2017}. Comparisons to other literature results require care, as our calculated values are the zero frequency limit of the frequency-dependent non-linear susceptibility. Nevertheless, we find generally good agreement between our calculations and experimental measurements~\cite{Wang2017, Kumar2013}. Our calculations indicate that the nonlinear susceptibilities of the h compounds change sign when going from S to Se to Te. The overall size of the tensor element decreases as the calculated band gap decreases, as was shown in other work~\cite{Wang2017}. 

The thermal and acoustic properties of our compounds come from calculations of the interatomic force constants including the dispersion correction~\cite{bulk_pike}. Our calculated phonon band structures, phonon density of states, and Raman spectra for the model compounds are shown in Fig.~\ref{three_phon}, with experimental Raman frequencies at $\Gamma$ from Refs. ~\cite{Lu2015, Ruppert2014, Scheuschner2015, Chang2014, Huo2015, Feng2015, Zhao2015, Zeng2013}. Here, the phonon density of states contributed by the transition metal atoms is given in blue and the chalcogen atoms in red. The nearly quadratic behavior near q $\rightarrow$ 0 indicates, as shown in Ref.~\cite{Andres2012}, that there is no internal strain within these materials after relaxation. The ZA mode is not purely quadratic as the rotational sum rule is not imposed explicitly in our calculations. We have investigated the effects of an increased q-mesh density on this ZA mode and are using a q mesh such that the frequencies of the ZA mode near q $\rightarrow$ 0 are converged to better than 0.1\%. Additionally, as reported in Refs.~\cite{Sohier2017} and ~\cite{Sohier2017a}, we find the LO and TO modes are strongly affected by the dielectric environment of the two-dimensional layer for q points near $\Gamma$. Our phonon band structures use the standard 3D Coulomb correction, and the LO-TO splitting is unrepresentative of a truly 2D system (Ref.~\cite{Sohier2017}). Our calculations of the Born effective charges compare well to other theoretical work~\cite{Sohier2017b}.

In some of these compounds, our phonon band structures show Kohn anomalies due to their small or zero band gaps~\cite{Kohn1959, Ataca2012, Dolui2016, Sugawara2016}.  When compared to their bulk counterparts~\cite{bulk_pike}, the Kohn anomalies appear at $M$ for both bulk and monolayer TiS$_2$ and TiSe$_2$. Whereas in the case of bulk TiTe$_2$ we find no Kohn anomaly at $M$ in the bulk but do find the Kohn anomaly at $M$ in the monolayer. This agrees with the fact that no charge density wave transition is observed in bulk TiTe$_2$ whereas one was observed in the monolayer~\cite{Chen2017}. Our calculations agree with other theory work~\cite{Duong2015} indicating the relative size of the Kohn anomaly is a strong function of the smearing used during the DFPT calculation. Here a smearing of 10K is used. Additionally, for the tc compounds our calculated phonon band structures show that they are stable at $\Gamma$,  which is consistent with the theoretical predictions in Ref.~\cite{Wolverson2016}, and are Raman active~\cite{Wolverson2016, Tongay2014, Feng2015, Zhao2015}. 

\section{Conclusions}

With distinctive crystal symmetries and dimensional effects, the transition metal dichalcogenides display a variety of properties that can be exploited for future device applications. Here our first-principles calculations of the monolayer TMDs reveal both significant similarities and differences between monolayer and bulk compounds, in terms of their electronic and vibrational properties. These are extremely significant when modeling the elastic/mechanical and dielectric properties of these materials. 

The dimensionality of the monolayer systems brings about changes to many of the physical and electric properties.  The most noticeable changes occur due to the switch from an indirect to a direct electronic band gap and, to a lesser extent, the appearance of finite nonlinear optical coefficients and piezoelectric coefficients in the hexagonal compounds. Likewise, the number and frequency of the Raman and Infrared active modes change in monolayers compared to bulk (for the h and t compounds). With the reduced dimensionality of the materials, we find the dynamic charge of the system is nearly unchanged for the h and t compounds but varies in magnitude and sign for the tc compounds, due to charge rearrangement and a reduced interlayer interaction. The accuracy of our calculations for many of the material properties when compared to experimental data instills confidence in our calculation methods: a comparison in Fig.~\ref{trends} shows a $\pm$3\% error in most compounds compared to experiment, and often better.

Finally, our calculations of monolayer TcS$_2$ and ReSe$_2$ indicate that these materials may not be stable, free-standing, monolayers compounds in the triclinic phase. When compared to our previous bulk calculations (Ref.~\citenum{bulk_pike}), it seems that stacking layers of these compounds may stabilize the triclinic phase, indicating the inter-layer interaction plays an under-appreciated role in stabilizing these compounds. With this in mind, our calculations of the vibrational and dielectric properties of these materials only require stability of the compound at $\Gamma$, which is clearly indicated in the phonon spectra. A determination of the low temperature phase diagram and symmetry class of these compounds is beyond the scope of this work. 

In conclusion, we have calculated the vibrational and dielectric properties of the most common transition metal dichalcogenides using the \textsc{Abinit} software package, the DFT-D3 van der Waals functional, GGA exchange-correlation and norm-conserving pseudo-potentials. Our calculations for the monolayer compounds are compared to bulk phases and experimental measurement, and show excellent agreement. We extract trends and outliers in the three symmetry classes. It is our hope that this information can be used as part of a broader effort to engineer heterostructures that combine the unique properties of the individual monolayer materials. 

\section*{Acknowledgments}
The authors gratefully acknowledge funding from the Belgian Fonds National de la Recherche Scientifique FNRS under grant numbers PDR T.1077.15-1/7 (N.A.P and M.J.V), PDR  T.0103.19 - ALPS (X.G and M.J.V) and for an FRIA Grant (B.V.T.). N.A.P would like to thank the Research Council of Norway through the Frinatek program for funding. M.J.V and A.D. acknowledge support from ULg and from the Communaut\'{e} Fran\c{c}aise de Belgique (ARC AIMED 15/19-09). Computational resources have been provided by the Consortium des Equipements de Calcul Intensif en F\'{e}d\'{e}ration Wallonie Bruxelles (CECI), funded by FRS-FNRS G.A. 2.5020.11; the Tier-1 supercomputer of the F\'{e}d\'{e}ration Wallonie-Bruxelles, funded by the Walloon Region under G.A. 1117545; and by PRACE-3IP DECI grants, on ARCHER and Salomon (ThermoSpin, ACEID, OPTOGEN, and INTERPHON 3IP G.A. FP7 RI-312763 and 13 G. A. 653838 of H2020).

\bibliography{tmd_source}

\end{document}